\documentclass[twocolumn,showpacs,preprintnumbers,amsmath,amssymb]{revtex4}
\usepackage{graphicx}
\begin{document}


\title{Temperature and magnetic field dependent tunneling spectroscopy of PtIr/Sr$_{0.9}$La$_{0.1}$CuO$_2$ point contact}

\author{L. Shan$^1$, H. Gao$^1$, Z.Y. Liu$^1$, Min-Seok Park$^2$, Kyung Hee Kim$^2$, C.U. Jung$^2$, Sung-Ik Lee$^2$}
\author{H. H. Wen$^1$} \email{hhwen@aphy.iphy.ac.cn}

\affiliation{$^1$ National Laboratory for Superconductivity,
Institute of Physics, Chinese Academy of Sciences, P.~O.~Box 603,
Beijing 100080, P.~R.~China}

\affiliation{$^2$ National Creative Research Initiative Center for
Superconductivity and Department of Physics, Pohang University of
Science and Technology, Pohang 790-784, Republic of Korea}

\date{\today}

\begin{abstract}
Low barrier quasiparticle tunneling spectroscopy on Sr$_{0.9}$La$_{0.1}$CuO$_2$ has been
studied with PtIr/Sr$_{0.9}$La$_{0.1}$CuO$_2$ point contacts at various temperatures and
magnetic fields. No zero-bias conductance peaks are observed. By fitting tunneling
conductance to Blonder-Tinkham-Klapwijk theory, the temperature dependent $s$-wave
superconducting gaps are obtained. The present results exclude the possibility of
$d$-wave symmetry for the pair gap in Sr$_{0.9}$La$_{0.1}$CuO$_2$, and do not support the
conventional phonon mediated pairing in this system either.

\end{abstract}


\maketitle

\section{introduction}
The symmetry of the superconducting order parameter is a crucial problem in revealing the
mechanism of superconductivity in cuprates. It is now generally agreed that hole-doped
($p$-type) high temperature superconducting cuprates have a $d_{x^2-y^2}$ pairing
symmetry in the underdoped and optimally doped region
\cite{TsueiCC2000,VanHarlingenDJ1995} while in the overdoped region a small component
such as $s$ or $d_{xy}$ appears in addition to the $d_{x^2-y^2}$ order parameter
\cite{YehNC2001,KhveshchenkoDV,SangiovanniG2003, VojtaM2000}. On the other hand, the
$d_{x^2-y^2}$ wave symmetry in the low and optimally electron-doped ($n$-type) curtates
has also been confirmed by the recent scanning SQUID \cite{TsueiCCprl2000}, tunneling
\cite{BiswasA2001}, ARPES \cite{ArmitageNP2001,SatoT2001}, penetration depth
\cite{KokalesJD2000,SkintaJA2002,ProzorovR2000}, and specific heat experiments
\cite{BalciH2002}. However, the $s$-wave symmetry is found to be dominant in the
overdoped $n$-type cuprates \cite{SkintaJA2002,BiswasA2002}, which is very different from
the case of $p$-type ones. Although there are significant differences between the
$p$-type and $n$-type cuprates both in structure and in superconductivity, it is of
general interest to see whether their behaviors can be explained within a unified
physical picture. Using the exchange of antiferromagnetic spin fluctuations as the
relevant pairing mechanism, Manske {\it et al.} \cite{ManskeD2000,ManskeD2003} have
theoretically demonstrated the existence of the pure $d_{x^2-y^2}$ wave symmetry in
underdoped $n$-type cuprates, they also successfully interpreted the well-known
experimental facts such as the smaller T$_c$ values of $n$-type cuprates than that of
$p$-type ones, and why the superconductivity of $n$-type cuprates only occurs for a
narrow doping range \cite{TakagiH1989,TokuraY1989,PengJL1997}. In Ref.\cite{ManskeD2000},
the electron-phonon coupling was introduced to explain the weakening of the $d_{x^2-y^2}$
pairing symmetry and the dominance of $s$-wave symmetry in overdoped $n$-type. That is,
there may be some kinds of interactions coexisting and competing with each other.
Recently observed nodeless gap in nearly whole doping range of
Pr$_{2-x}$Ce$_x$CuO$_{4-\delta}$ ($0.115\leq x \leq 0.152$) superconducting films
\cite{KimMS2003} also implicated that some order parameter may prevail over the others at
some given conditions. Therefore, in order to understand the electronic phase diagram and
the superconduting mechanism of $n$-type cuprates in whole doping range, one of the
essential tasks is to confirm the possible pairing symmetry other than $d_{x^2-y^2}$ type
and investigate its properties in detail.

The infinite-layer system Sr$_{1-x}$La$_{x}$CuO$_2$ is the simplest form of $n$-type
cuprates. The recently synthesized single-phase samples of Sr$_{0.9}$La$_{0.1}$CuO$_2$
have nearly $100\%$ superconducting volume \cite{JungCU2002}, enabling reliable studies
on its superconductivity. This system is of interest because of its three-dimensional
superconductivity \cite{KimMS2001} and it is expected to be a good candidate showing
isotropic order paremeter. Recently, its $s$-wave superconductivity has been proposed by
scanning tunneling microscopy (STM) \cite{ChenCT2002} and specific heat \cite{LiuZY2003}
measurements separately. However, more detailed information is needed to clarify its
pairing symmetry and superconducting mechanism.

In this work we study the quasiparticle tunneling spectroscopy of
PtIr/Sr$_{0.9}$La$_{0.1}$CuO$_2$ point contacts. The $s$-wave pairing symmetry is
demonstrated both by fitting the measured spectra to Blonder-Tinkham-Klapwijk (BTK)
theory and by investigating the field induced quasipartical states. Meanwhile, the
determined temperature dependence of superconducting energy gap $\Delta(T)$ is obviously
different from that of the conventional electron-phonon coupling superconductors.

\section{experimental method}
\begin{figure}[top]
\includegraphics[scale=1.2]{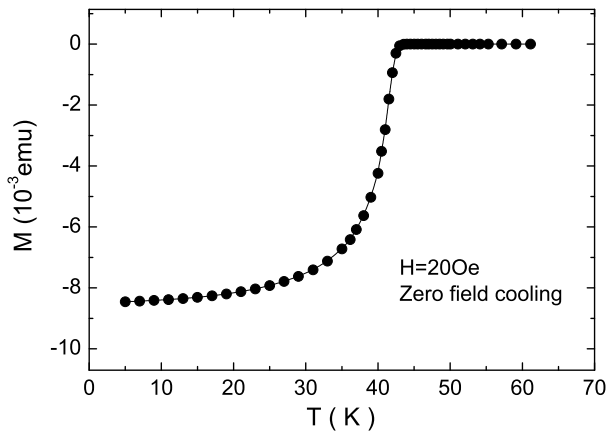}
\caption{\label{fig:fig1} Superconducting transitions measured by Dc magnetization. }
\end{figure}

The sample studied in this work is high-density granular material of
Sr$_{0.9}$La$_{0.1}$CuO$_2$ \cite{JungCU2002}, which is the same as that reported in Refs
\cite{ChenCT2002,LiuZY2003}. X-ray diffraction confirmed the single-phase nature of the
sample and revealed random grain orientation \cite{JungCU2002}. The sample was cut into a
slab with dimensions of about $2\times2\times0.5mm^3$. As shown in Fig.~\ref{fig:fig1},
the superconducting transition temperature T$_c$ is about 43K as characterized by DC
magnetization measurement. The sample surface was polished by fine metallographic
grinding papers and then milled by Ar iron for 20 minutes. The PtIr tips were prepared by
electrochemical etching in CaCl$_2$ solution using Pt$_{0.9}$Ir$_{0.1}$ wire with a
diameter of 0.25 mm. The approaching of the tips was controlled by a differential screw.
The point contact insert was set in the sample chamber of an Oxford cryogenic system
Maglab. Typical four-terminal and lock-in technique were used to measure the differential
resistance $dV/dI$ vs $V$ of the point contacts. Then the $dV/dI-V$ curves were converted
into the dynamical conductance $dI/dV-V$ (or $\sigma-V$) curves.

\begin{figure}
\includegraphics[scale=1.2]{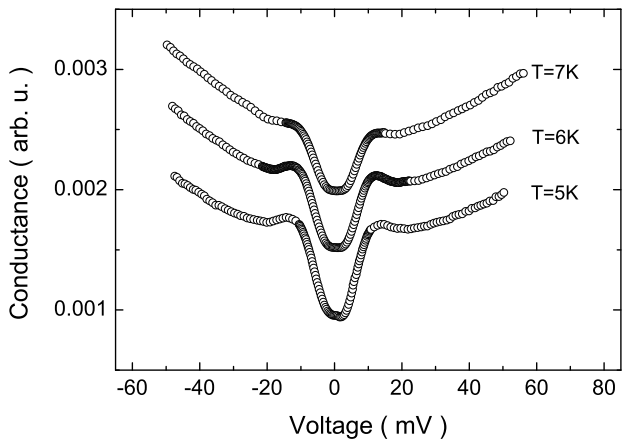}
\caption{\label{fig:fig2} Typical conductance spectra of low-barrier tunneling
measurements on PtIr/Sr$_{0.9}$La$_{0.1}$CuO$_2$ point contacts.}
\end{figure}

\begin{figure}
\includegraphics[scale=1.2]{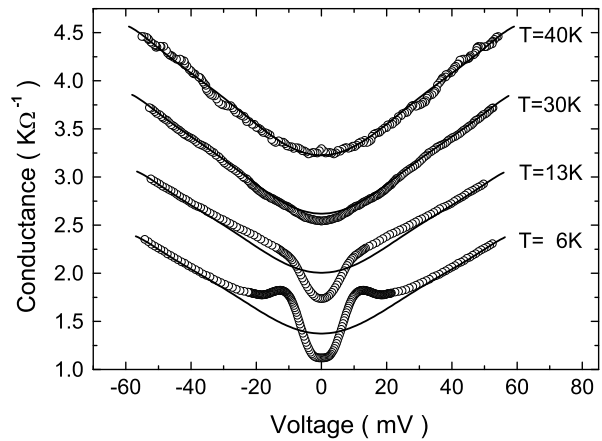}
\caption{\label{fig:fig3} An illustration of constructing normal conductance background
for various temperatures according to the functional form of the spectra around T$_c$.}
\end{figure}

\section{results and discussion}
\begin{figure}
\includegraphics[scale=1.2]{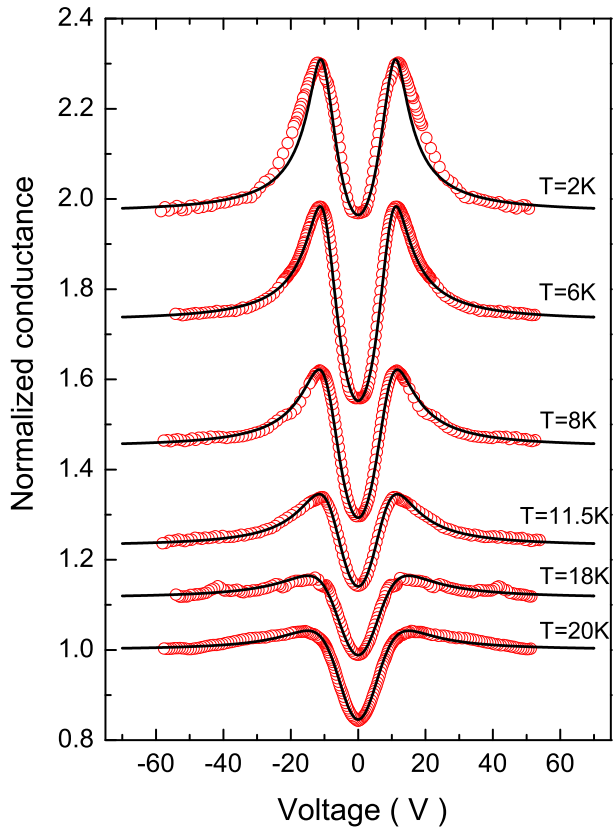}
\caption{\label{fig:fig4} Theoretically fitting the normalized conductance for various
temperatures. The normalized spectra are denoted by open circles. The solid lines
represent the fits to the modified BTK theory. All curves except the bottom curve have
been shifted upwards for clarity.}
\end{figure}

Fig.~\ref{fig:fig2} shows the typical raw data of the conductance ($\sigma$) of
PtIr/Sr$_{0.9}$La$_{0.1}$CuO$_2$ point contact. In the following analysis, in order to
reduce the uncertainty in the fitting process, we have averaged the positive and negative
bias parts for all the measured spectra as shown in Fig.~\ref{fig:fig3}. It is noted that
all the characteristics of superconductivity disappear completely at about 40K, therefore
we construct the background of each spectral curve according to the functional form of
the normal-state conductance (at T=40K), as denoted by solid lines presented in
Fig.~\ref{fig:fig3}. All spectra are normalized by the corresponding background before
fitting.  The typical normalized spectra are presented in Fig.~\ref{fig:fig4}, denoted by
open circles. It is found that all the curves for lower temperature can be well fitted
with the generalized BTK theory \cite{BlonderGE1982} with BCS density of states (DOS), in
which the quasiparticle energy $E$ is replaced by $E+i\Gamma$, where $\Gamma$ is the
smearing parameter characterizing the finite lifetime of the quasiparticles
\cite{PlecenA1994}. For the temperature higher than 20K, the spectra are badly smeared
due to the thermal effect and surface roughness, therefore, it is difficult to obtain the
credible fitting parameters. Another significant result of our experiments is that no
zero-bias conductance peak (ZBCP) is found in the spectra for any grain orientations.
This is consistent with the results of STM \cite{ChenCT2002} and has been regarded as a
proof of non-$d$-wave pairing symmetry.

\begin{figure}
\includegraphics[scale=1.2]{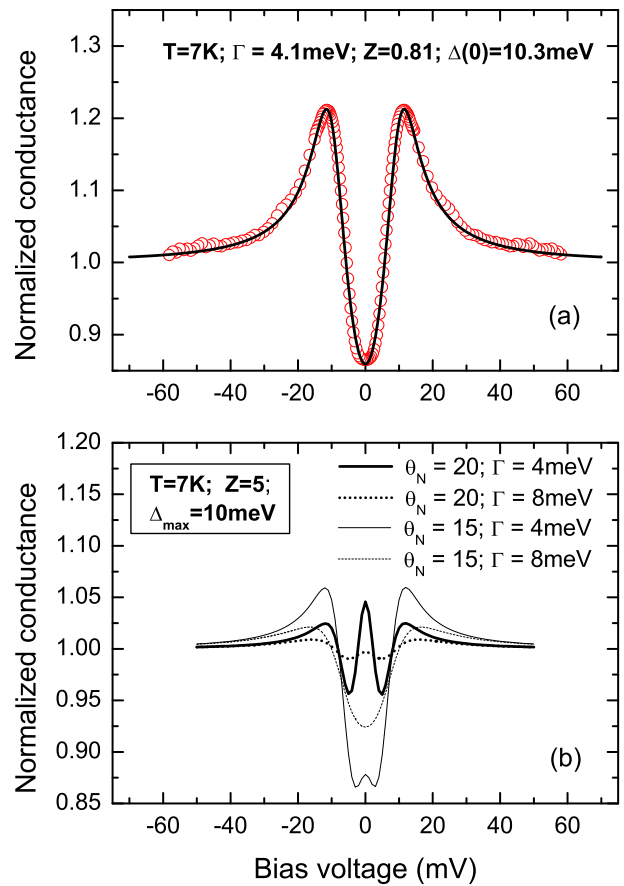}
\caption{\label{fig:fig5} (a) Adopted parameters for a typical s-wave fit to modified BTK
theory. (b) Theoretical calculations for the spectra of a superconductor with
$d_{x^2-y^2}$ pairing symmetry.}
\end{figure}

However, in some tunneling experiments, no ZBCP was observed even in the superconductors
with $d$-wave symmetry \cite{BiswasA2002,ApriliM1998}. There are two possible reasons.
One is the effect of disorder and the other is thermal smearing. We have examined these
two effects in our data. As shown in Fig.2 of Ref.\cite{ApriliM1998}, when the ZBCP is
suppressed by an increase of disorder in the material, the whole spectral curve becomes
unfeatured, which is not the case in our work, i.e., although there is no ZBCP, the other
characteristics of superconductivity are prominent. Furthermore, the smearing effect is
quantitatively investigated by introducing a smearing factor $\Gamma$ into the spectral
calculations for $d_{x^2-y^2}$ symmetry. In such calculations, we adopt the same
potential model and formula presented in Ref.\cite{KashiwayaS1996}. The theoretical
simulations of the spectra for two arbitrarily selected injection angles ($\theta_N=15^o$
and $\theta_N=20^o$) are presented in Fig.~\ref{fig:fig5}(b), a typical modified BTK
fitting is shown in Fig.~\ref{fig:fig5}(a) as a comparison. It is found that although the
smearing factor $\Gamma$ is as large as $40\%$ of the maximal gap $\Delta_{max}$, the
ZBCP is still prominent. Even the $\Gamma$-value is enhanced to 8meV which has been
comparable to $\Delta_{max}$, the ZBCP still exists for $\theta_N=20^o$. Moreover, the
depression of ZBCP is accompanied by the remarkable weakening of all the spectral
features. It should also be pointed out that the ZBCP will become more pronounced when
$\theta_N$ approaches to $45^o$. Considering the random orientations of the crystal
grains, if the pairing symmetry is $d_{x^2-y^2}$ type, the measured spectra may have
remarkable difference from one position to the other and the ZBCP should be easily
observed at many grains. Therefore, a fully gapped fermi surface should be responsible
for the non-existence of ZBCP in Sr$_{0.9}$La$_{0.1}$CuO$_2$.

\begin{figure}[top]
\includegraphics[scale=1.2]{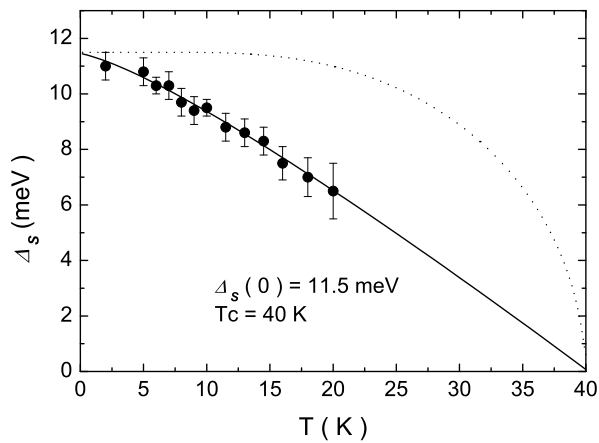}
\caption{\label{fig:fig6} The experimental energy gap (solid circles) obtained from the
fits described in Fig.~\ref{fig:fig4}. The dotted line is the prediction of conventional
electron-phonon coupling mechanism. The solid line is the guide to eyes.}
\end{figure}

In Fig.~\ref{fig:fig6}, we present the temperature dependence of superconducting gap
determined by fitting the normalized conductance spectra to modified BTK theory. The
theoretical prediction of electron-phonon coupling mechanism is also presented as a
comparison. Although the $s$-wave pairing symmetry and the BCS-type quasiparticle
dispersion ($E_{\pm}(k)=\pm\sqrt{\varepsilon(k)^2+\Delta^2}$) are accepted in our
fitting, the obtained superconducting gap $\Delta(T)$ obviously deviates the conventional
electron-phonon coupling mechanism. The BCS-type quasiparticle dispersion has recently
been observed in Bi-2212 by STM \cite{HoffmanJE2002} and in Bi-2223 by ARPES
\cite{MatsuiH2003}. The good agreement of theoretical fitting and the experimental
results in our work indicates that the infinite-layer $n$-type cuprate
Sr$_{0.9}$La$_{0.1}$CuO$_2$ also favors such quasiparticle dispersion.

\begin{figure}[top]
\includegraphics[scale=1.2]{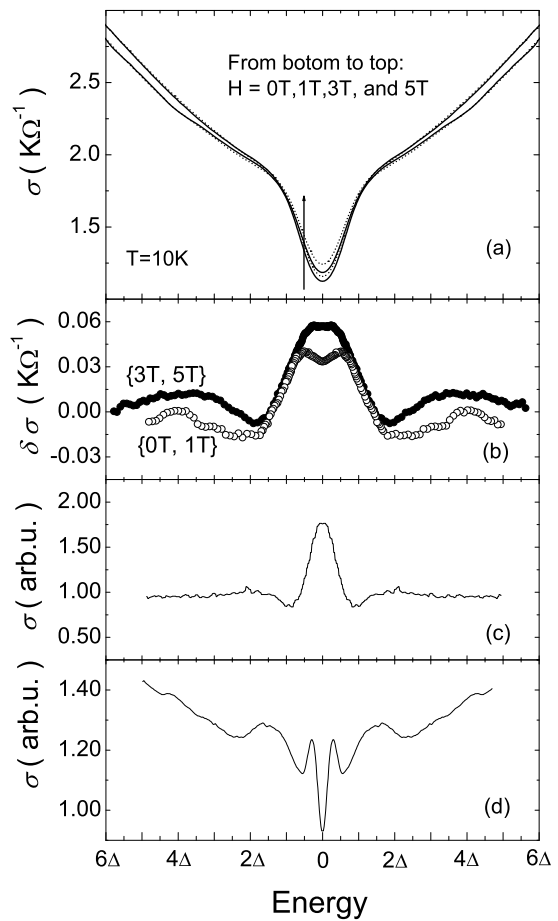}
\caption{\label{fig:fig7} (a) The typical magnetic field dependence of the tunneling
spectra of PtIr/Sr$_{0.9}$La$_{0.1}$CuO$_2$ point contacts measured at 10K. The applied
field is 5T, 3T, 1T, and 0T from top to bottom. (b) Difference between the conductance of
1T and 0T (open circles) and that of 5T and 3T (solid circles). (c) and (d) are the
typical spectrum around vortex cores in conventional superconductor NbSe$_2$
\cite{HessHF1989} and $p$-type cuprate YBa$_2$Cu$_3$O$_7$ \cite{Maggio-AprileI1995},
respectively. The abscissa of each figure has been normalized according to the energy gap
of the corresponding sample.}
\end{figure}

In order to get more information of superconductivity in Sr$_{0.9}$La$_{0.1}$CuO$_2$, we
have measured the response of the conductance spectra to magnetic field at various
temperture. In such measurements, we have pressed the tip tightly on the sample in order
to make the interface large enough so as to comprise enough vortices. Therefore, any
spectral variation caused by the change of vortex density can be detected when the
applied magnetic field is altered. If the applied field does not change the configuration
of the point contact, the background should be identical for the spectra measured at
different fields. From Fig.~\ref{fig:fig7}(a), we can see the superposition of the high
bias spectra for 0T and 1T, and this is also the case for 3T and 5T. However, there is a
clear difference between the background conductance of \{0T, 1T\} and that of \{3T, 5T\},
indicating that a slight shift of the tip has occurred. As a consequence, the spectral
shape is transformed not only by magnetic field but also by the potential barrier between
the tip and sample. Therefore, in order to investigate the field induced conductance, in
the following analysis we only study the spectra with the same background conductance.
Fig.~\ref{fig:fig7}(b) shows the difference between the conductance of 0T and 1T by open
circles, and that between 3T and 5T by solid circles, these are determined by subtracting
lower field data from higher field data.

\begin{figure}[top]
\includegraphics[scale=1.2]{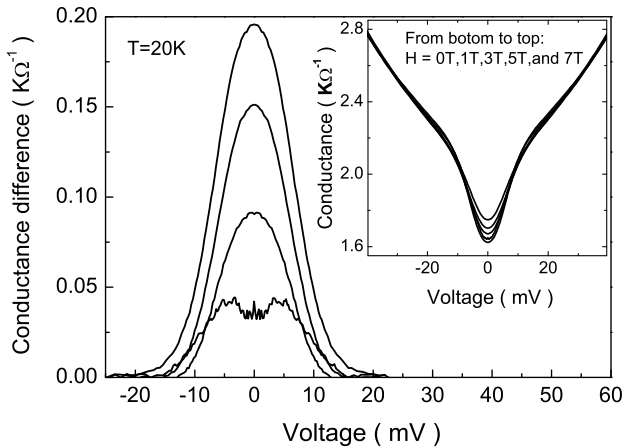}
\caption{\label{fig:fig8} Magnetic field induced conductance calculated from the raw data
shown in the inset. The spectra are measured at 20K and all the curves presented here
have the same background. }
\end{figure}

It is found that the field induced conductance (reflecting the vortex states) has two
common features for various temperatures. Firstly, a ZBCP occurs at low energy scale
around fermi energy and enhances with increasing magnetic field; Secondly, at low field
(e.g. 1T), the ZBCP splitting is observed. These two features are opposite to that of
vortex states in $p$-type cuprates \cite{PanSH2000,Maggio-AprileI1995}, which has no ZBCP
due to the appearance of field induced orders such as antiferromagnetism
\cite{TsuchiuraH2003} and d-density wave \cite{MaskaMM2003}. Furthermore, the ZBCP
splitting has not been observed around the vortex core in conventional type-II
superconductors \cite{HessHF1989}. However, these features are consistent with the
theoretical results for $s$-wave as reported by Wang {\it et al.} \cite{WangY1995}. In
their calculations, the microscopic mean-field theory was adopted and the structure of
CuO$_2$ layers was considered. The ``$s$-wave" model was supposed to have on-site
attraction. According to this model, a clear ZBCP is obtained at the center of the vortex
cores and the split occurs when the distance between the vortices is large enough (i.e.,
at low field). Although this model does not consider the coupling between CuO$_2$ layers,
its outcome is qualitatively consistent with our experimental results.

\begin{figure}[top]
\includegraphics[scale=1.15]{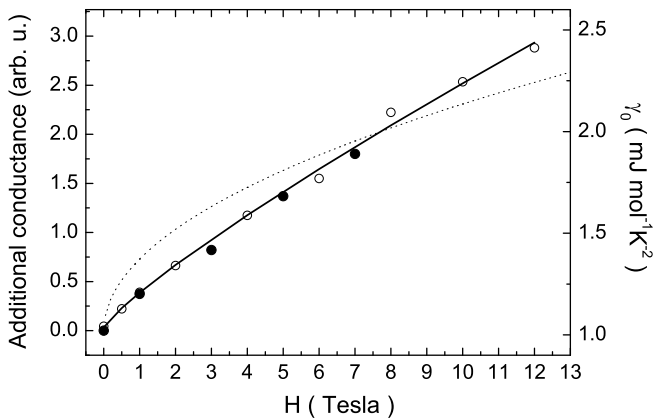}
\caption{\label{fig:fig9} The comparison between the field dependence of the additional
conductance in this work (solid circles) and that of low energy quasiparticle DOS
obtained from specific heat measurements (open circles). The dotted line is the
prediction for clean d-wave superconductors and the solid line has the form of $y\approx
H^{0.85}$. }
\end{figure}

In the inset of Fig.~\ref{fig:fig8}, we present a set of field dependent spectra with the
same background. As shown in the main frame of Fig.~\ref{fig:fig8}, the field induced
conductance (reflecting vortex states) has been determined by subtracting the zero-field
spectrum from that measured in various fields. The ZBCP splitting still can be recognized
for the lowest field (1T) although the temperature is as high as 20K. It is known that
the field induced quasiparticle density of states (DOS) mostly comes from the exterior of
the vortex cores for $d_{x^2-y^2}$ wave symmetry, while for $s$-wave symmetry the
contribution from the interior of the cores is dominant. Since the interface between the
tip and sample is large enough as we have mentioned, we have measured the total field
induced DOS both the outside and the inside of the vortex cores. We have integrated the
low energy additional conductance in a range of $\pm 5$meV, as shown in
Fig.~\ref{fig:fig9} (Two other different integral ranges of $\pm 10$meV and $\pm 30$meV
have also been selected and the obtained rules are the same as reported here). As a
comparison, we present the magnetic field dependent low energy DOS of
Sr$_{0.9}$La$_{0.1}$CuO$_2$ determined by specific heat measurements \cite{LiuZY2003}.
The results reported here are in good agreement with the specific heat measurements,
i.e., the field induced low energy DOS can be described as $y\approx H^m$ with $m=0.85$.
It is well known that the exponent $m$ should be 0.5 for clean $d$-wave superconductors
and 1 for $s$-wave superconductors in simple limit. However, $m$ may deviate these two
values in the case of dirty $d$-wave \cite{KubertC1998} or in the case of $s$-wave
considering the vortex lattice effect \cite{IchiokaM1999}. Ref.\cite{LiuZY2003} has ruled
out the possibility of dirty d-wave and adopted that Sr$_{0.9}$La$_{0.1}$CuO$_2$ is a
s-wave superconductor, which is in good agreement with our forementioned conclusions. It
should be pointed out that our result in Fig.~\ref{fig:fig9} is derived from the data
measured at 20K which is higher than that ($<7K$) in Ref.\cite{LiuZY2003}. However, in
the case of $s$-wave symmetry, the field dependence of additional DOS is mainly
determined by the density of vortices instead of temperature. Considering the much lower
magnetic field than the upper critical field and T$_c\approx40$K$>>20$K in our
investigations, the conclusion presented in Fig.~\ref{fig:fig9} is reasonable and
self-consistent.

\section{concluding remarks}
In summary, we have performed low-barrier quasiparticle tunneling measurements on the
PtIr/Sr$_{0.9}$La$_{0.1}$CuO$_2$ point contacts in detail. No ZBCP is observed on the
randomly oriented crystal grains. All the normalized conductance spectra can be well
fitted in the frame of BTK theory, providing an impelling evidence for the BSC-type
s-wave superconductivity in Sr$_{0.9}$La$_{0.1}$CuO$_2$. This is also proved by
investigating the magnetic field induced vortex states. However, it is noted that the
temperature dependence of superconducting gap deviates from the prediction of
conventional electron-phonon coupling mechanism, and the field induced ZBCP is splitted
at low field. These features can not be interpreted in the frame of conventional
superconducting mechanism, therefore, cautious must be taken to associate the
experimentally observed $s$-wave superconductivity in overdoped $n$-type cuprates with
the electron-phonon mechanism.

\begin{acknowledgments}
This work is supported by the National Science Foundation of China (NSFC 19825111,
10074078), the Ministry of Science and Technology of China ( project: NKBRSF-G1999064602
), and Chinese Academy of Sciences with the Knowledge Innovation Project. The work in
Pohang University is supported by the Ministry of Science and Technology of Korea through
the Creative Initiative Program.
\end{acknowledgments}


\end{document}